\documentclass[preprint,showpacs,showkeys,aps]{revtex4}

\usepackage{graphicx}

\newcommand{\myfigure}[3]{
        \begin{figure}
        \centerline{
        \includegraphics{#1.eps}}
        \caption{#2}
        \label{#3}
        \end{figure}
}

\newcommand{\bq}{\begin{equation}}
\newcommand{\eq}{\end{equation}}
\newcommand{\bqn}{\begin{eqnarray}}
\newcommand{\eqn}{\end{eqnarray}}
\newcommand{\nb}{\nonumber}
\newcommand{\lb}{\label}

\begin{document}

\title{Dressing a Naked Singularity: an Example}

\author{C. F. C. Brandt}
\email{fredcharret@yahoo.com.br}
\affiliation{Departamento de
F\'{\i}sica Te\' orica, Universidade do Estado do Rio de Janeiro
(UERJ), Rua S\~ ao Francisco Xavier $524$, Maracan\~ a, CEP
20550-013, Rio de Janeiro, RJ, Brazil}

\author{R. Chan}
\email{chan@on.br}
\affiliation{Coordena\c c\~ao de Astronomia e
Astrof\'{\i}sica, Observat\'orio Nacional, Rua General Jos\'e
Cristino 77, S\~ao Crist\'ov\~ao, CEP 20921-400, Rio de Janeiro, RJ,
Brazil}

\author{M. F. A. da Silva}
\email{mfasnic@gmail.com}
\affiliation{Departamento de F\' {\i}sica
Te\' orica, Universidade do Estado do Rio de Janeiro, Rua S\~ ao
Francisco Xavier $524$, Maracan\~ a, CEP 20550-013, Rio de Janeiro,
RJ, Brazil}

\author{Jaime F. Villas da Rocha}
\email{jfvroch@pq.cnpq.br}
\affiliation{Universidade Federal do Estado do Rio de Janeiro,
Instituto de Bioci\^encias,
Departamento de Ci\^encias Naturais, Av. Pasteur 458, Urca, 
CEP 22290-240, Rio de Janeiro, RJ, Brazil}

\date{\today}

\label{firstpage}

\begin{abstract}

Considering the evolution of a perfect fluid with self-similarity of
the second kind, we have found that an initial naked singularity 
can be trapped by an event horizon due to collapsing matter.
The fluid moves along time-like geodesics with a
self-similar parameter $\alpha = -3$.  Since the metric obtained is
not asymptotically flat, we match the spacetime of the fluid with a
Schwarzschild spacetime.
All the energy conditions are fulfilled until the naked
singularity.
\end{abstract}

\keywords{gravitational collapse, black hole, naked singularity}

\pacs{04.20.Jb, 04.40.Dg, 97.10.-q}

\maketitle

\section{Introduction}
One of the most important problem in the relativistic astrophysics
today is the final fate of a massive star,
which enters the state of an endless gravitational collapse once it has
exhausted its nuclear fuel. What will be the end state of such a
continual collapse which is entirely dominated by the force of gravity?
The conjecture that such a collapse, under physically realistic
conditions, must end into the formation of a black hole is called the
Cosmic Censorship Conjecture (CCC), introduced by Penrose and Hawking \cite{rev1}. 
Despite numerous attempts over the
past three decades, such a conjecture remains unproved and continues to
be a major unsolved problem, lying at the foundation of the theory
and applications in black hole physics.

A considerable debate has continued on the validity or otherwise of the CCC, 
which effectively states that any
singularities that arise from gravitational collapse from a regular
initial data must not be visible to far away observers in the spacetime
and are always hidden within black holes.
Such an assumption has been used extensively and is fundamental to the
theory as well as applications of black hole physics.
On the other hand, if a naked singularity results as collapse end
state, it is no longer necessarily
covered by the event horizon and could communicate, in principle,
with outside observers. Such a scenario is of physical interest because
a naked singularity may have theoretical and observational properties
quite different from a black hole end state, and communications from
extreme strong gravity regions dominated by quantum gravity
may be possible.

Although there is no satisfactory proof or mathematical formulation
of CCC available despite many efforts, there are many examples of
dynamical collapse models available which lead to a black hole or
a naked singularity as the collapse end state, depending on the
nature of the initial data (see e.g.  \cite{rev1}-\cite{rev8}
and references therein).
In particular, pioneering analytic models by  Christodoulou,
and Newman, and numerical work by Eardley and Smarr \cite{dust1}-\cite{dust3},
established the existence of shell-focusing
naked singularities as the end state of a continual collapse,
where the physical radius of all collapsing shells vanishes.
In these models the matter form was
taken to be marginally bound dust, assuming that the initial
data functions are smooth and even profiles.
Newman generalized these models for non-marginally bound class,
thus covering the entire class of dust collapse solutions.
A general treatment for dust collapse with generic initial data
was developed in \cite{jd}.

Here we present an example an initial naked singularity can evolve to
a black hole by the collapse of matter.  We consider a perfect
fluid which moves along time-like geodesics with a
self-similar parameter $\alpha = -3$ and satisfies all the energy
conditions.

The paper is organized as follows. In Section
II we present the Einstein field equations. In Section III we present 
the exact solution that represents a perfect fluid moving
along time-like geodesics for the self-similar parameter $\alpha = -3$.
 The ingoing, outgoing null
congruence scalar expansions and the energy conditions \cite{HE73} are
 analyzed.
In section IV we match the fluid with Schwarzschild space time to
obtain an asymptotically flat condition garanting the final structure
 of a black hole or a naked singularity.
 Finally, in Section V we present the conclusions.

\section{The Field Equations}

The general metric of spacetimes with spherical symmetry  can be
cast in the form,
\bq \lb{genmetric} ds^2 = r_1^2 \left[e^{2
\Phi(t,r)}dt^2 - e^{2 \Psi(t,r)} dr^2 - r^2 S^2(t,r)
d\Omega^2\right],
\eq
 where $d\Omega^2 =  d\theta^2 + \sin^2\theta
d\phi^2 $, and $r_{1}$ is a constant with the dimension of length.
Then, we can see that the coordinates $t, r, \theta$ and $\phi$, as
well as the functions $\Phi, \Psi$ and $S$ are all dimensionless.

Self-similar solutions of the second kind  are given by
\cite{Brandt2003}
 \bq
\lb{seckind}
 \Phi(t,r) = \Phi(x),\;\;\; \Psi(t,r) = \Psi(x),\;\;\;
S(t,r) = S(x), \eq where \bq \lb{ss} x \equiv \ln \left[
\frac{r}{(-t)^{1/\alpha}} \right], \eq and $\alpha$ is a
dimensionless constant. The  energy-momentum tensor of the perfect
fluid is written in the form \bq \lb{emtgen} T_{\mu \nu} = (\rho +
p) u_{\mu} u_{\nu}
 - p g_{\mu \nu}
 \eq
 where $u^{\mu}$
denotes the four-velocity of the fluid, while $\rho$ and $p$ stands
for the energy density and the pressure of the fluid. Then, we can
see that $\rho $ is the energy density of the fluid measured by
observers comoving with the fluid. In the comoving coordinates, we
have
\bqn
 \lb{velocity}
  u_{\mu} &=& e^{\Phi(x)} \delta_{\mu}^t  .
 \eqn

Defining \bq
\lb{defy}
 y \equiv \frac{\dot S}{S},
 \eq
 where the symbol dot
over the variable denotes differentiation with respect to $x$. The
non-null  components of the Einstein tensor for the metric given by 
equation (\ref{genmetric}) with equations (\ref{seckind})-(\ref{defy})
can be written as
 \bqn
\lb{C.4a} G_{tt} &=& - \frac{1}{r^{2}}e^{2(\Phi -\Psi)} \left[2\dot
y + y(3y +
4) + 1 - 2(1+y)\dot \Psi - S^{-2}e^{2\Psi}\right]\nb\\
& & + \frac{1}{\alpha^2 t^{2}}(2\dot \Psi + y)y,\\
\lb{C.4b} G_{tr} &=&  \frac{2}{\alpha t r}
\left[\dot y + (1+y)(y - \dot \Psi) - y\dot \Phi\right],\\
\lb{C.4c} G_{rr} &=& \frac{1}{r^{2}}\left[
2(1+y)\dot \Phi + (1 + y)^{2}  - S^{-2}e^{2\Psi}\right]\nb\\
& & - \frac{1}{\alpha^2 t^{2}}e^{2(\Psi - \Phi)}
\left[2\dot y + y\left(3y - 2\dot \Phi + 2\alpha\right)\right],\\
\lb{C.4d} G_{\theta\theta} &=&  S^{2}e^{-2\Psi}\left[\ddot \Phi +
\dot y + \dot \Phi\left(\dot \Phi - \dot \Psi + y\right) +
\left(1 + y\right)\left(y - \dot \Psi\right)\right]\nb\\
& &  - \frac{r^{2}S^{2}}{\alpha^2 t^{2}}e^{-2\Phi}\left[\ddot \Psi +
\dot y + y^{2} - \left(\dot \Psi + y\right)\left(\dot \Phi - \dot
\Psi - \alpha \right)\right], \eqn where in writing the above
expressions we have set $r_{1} = 1$.

In the next section we have solved the Einstein's equations, 
$G_{\mu\nu}=T_{\mu\nu}$.

\section{Geodesic Model}

We study now the solution of the perfect fluid with self-similarity
in a geodesic model, that is, a situation in which the acceleration
$\dot{\Phi} = 0$, and in particular we made $\Phi = 0$.  Thus, we
have
 \bqn
 \lb{einstein1}
G_{tt} &=& - \frac{1}{r^{2}}e^{-2 \Psi} \left[2\dot y + y(3y +
4) + 1 - 2(1+y)\dot \Psi - S^{-2}e^{2\Psi}\right]\nb\\
& & + \frac{1}{\alpha^2 t^{2}}(2\dot \Psi + y)y,\\
\lb{einstein2} G_{tr} &=&  \frac{2}{\alpha t r}
\left[\dot y + (1+y)(y - \dot \Psi) \right],\\
\lb{einstein3} G_{rr} &=& \frac{1}{r^{2}}\left[
 (1 + y)^{2}  - S^{-2}e^{2\Psi}\right]\nb\\
& & - \frac{1}{\alpha^2 t^{2}}e^{2 \Psi}
\left[2\dot y + y\left(3y  + 2\alpha\right)\right],\\
\lb{einstein4} G_{\theta\theta} &=&  S^{2}e^{-2\Psi}\left[ \dot y +
\left(1 + y\right)\left(y - \dot \Psi\right)\right]\nb\\
& &  - \frac{r^{2}S^{2}}{\alpha^2 t^{2}}\left[\ddot \Psi + \dot y +
y^{2} - \left(\dot \Psi + y\right)\left( - \dot \Psi - \alpha
\right)\right]
\eqn

The Einstein field equations yield
   \bqn 
   \lb{einstein5}
G_{tr} &=&  \frac{2}{\alpha t
r} \left[\dot y + (1+y)(y - \dot \Psi) \right]=0 ,
\lb{Gtr1}
 \eqn
and
\bqn
 \lb{einstein2a}
  \rho &=& - \frac{1}{r^{2}}e^{-2 \Psi} \left[2\dot y + y(3y +
4) + 1 - 2(1+y)\dot \Psi - S^{-2}e^{2\Psi}\right]\nb\\
& & + \frac{1}{\alpha^2 t^{2}}(2\dot \Psi + y)y  . 
\eqn

As we study the perfect fluid in this work, then we should have 
$G^r_r = G^{\theta}_{\theta} $, that furnishes
\bqn
   p &=&  \frac{e^{-2\Psi}}{r^{2}}\left[
(1 + y)^{2}  - S^{-2}e^{2\Psi} \right] - \frac{1}{\alpha^2 t^{2}} \left[2\dot
 y + y\left(3y  +
2\alpha\right)\right]     \nb\\
     &=&   \frac{e^{-2\Psi}}{r^2}\left[ \dot
y + 
\left(1 + y\right)\left(y - \dot \Psi\right)\right],\nb\\
& &  - \frac{1}{\alpha^2 t^{2}}\left[\ddot \Psi + \dot y + y^{2} -
\left(\dot \Psi + y\right)\left( - \dot \Psi - \alpha \right)\right]
.
\lb{press}
   \eqn
Thus, we can obtain from (\ref{press}) the following equations
\bq
(1 + y)^{2}  - S^{-2}e^{2\Psi} =
\dot y + \left(1 + y\right)\left(y - \dot \Psi\right) = 0,
\lb{pressa}
\eq
because of equation (\ref{einstein5}) and
\bq
2\dot y + y\left(3y  + 2\alpha\right) =
\ddot \Psi + \dot y + y^{2} - \left(\dot \Psi + y\right)\left( - \dot \Psi - \alpha \right).
\lb{pressb}
\eq

Substituting the equation (\ref{Gtr1}) into (\ref{pressb})
we get the following differential equation 
\bqn
\lb{diferentialequation}
  \ddot{y} + 3y \dot{y} + \alpha\dot{y} = 0, 
	\eqn
for $y \neq -1$ and $\alpha=-3$, we have the particular solution 
      \bqn
      \lb{solperfectfluid}
      S(X) &=& S_0 X^{2/3} e^{X} ,  \nb\\
       e^{\Psi(X)} &=& S_0 e^X \left[ \frac{6X+2}{3X^{1/3}} \right]
      \eqn
       with $S_0 > 0$ and $X \equiv ln( r(-t)^{1/3} )$. The resulting
 metric 
can be written as
      \bqn
       \lb{metricX}
       ds^2 = dt^2 - S_0^2 X^{4/3} e^{2X} \left[ 4 \left( 1 +
 \frac{1}{3X}\right)^2 dr^2 -r^2 d\Omega^2 \right]   .
       \eqn

         The geometric radius is given by
          \bqn
          \lb{geometricradius}
           R = r S = S_0 r^2 (-t)^{1/3} [ln(r(-t)^{1/3})]^{2/3}.
              \eqn
         Note that the first time derivative of the geometric radius
         is negative, indicating that the fluid is collapsing.

        Considering the solution (\ref{solperfectfluid}) in
 (\ref{einstein2a}) we
        obtain the energy density and pressure, given by
        \bq
        \lb{endensityperfluid}
   \rho = \frac{1}{3t^2} \left[ \frac{ 2 + 3X}{1+ 3 X} \right]
   \left[\frac{1}{X} +1 \right],
       \eq
and
        \bq
        \lb{pressureperfluid}
        p =  \frac{1}{3t^2}
        .
         \eq

In order to analyze the physical singularities we get the Kretschmann scalar,
given by
\bq
K=R_{\alpha\beta\gamma\lambda}R^{\alpha\beta\gamma\lambda}=\frac{4}{2187} 
\frac{3645X^6+3618X^5+5265X^4+2268X^3+441X^2+48X+16}
{t^4 X^4 (3X+1)^2}. 
\label{K}
\eq

       Observing the energy density (\ref{endensityperfluid}),  we
 identify three
       singularities, which are $t=0$, $X = 0$, $X = - 1/3$. The singularity 
$X=0$ is the outest, which in terms of the $t$
 and
$r$ coordinates we have (see figure 1) 
\myfigure{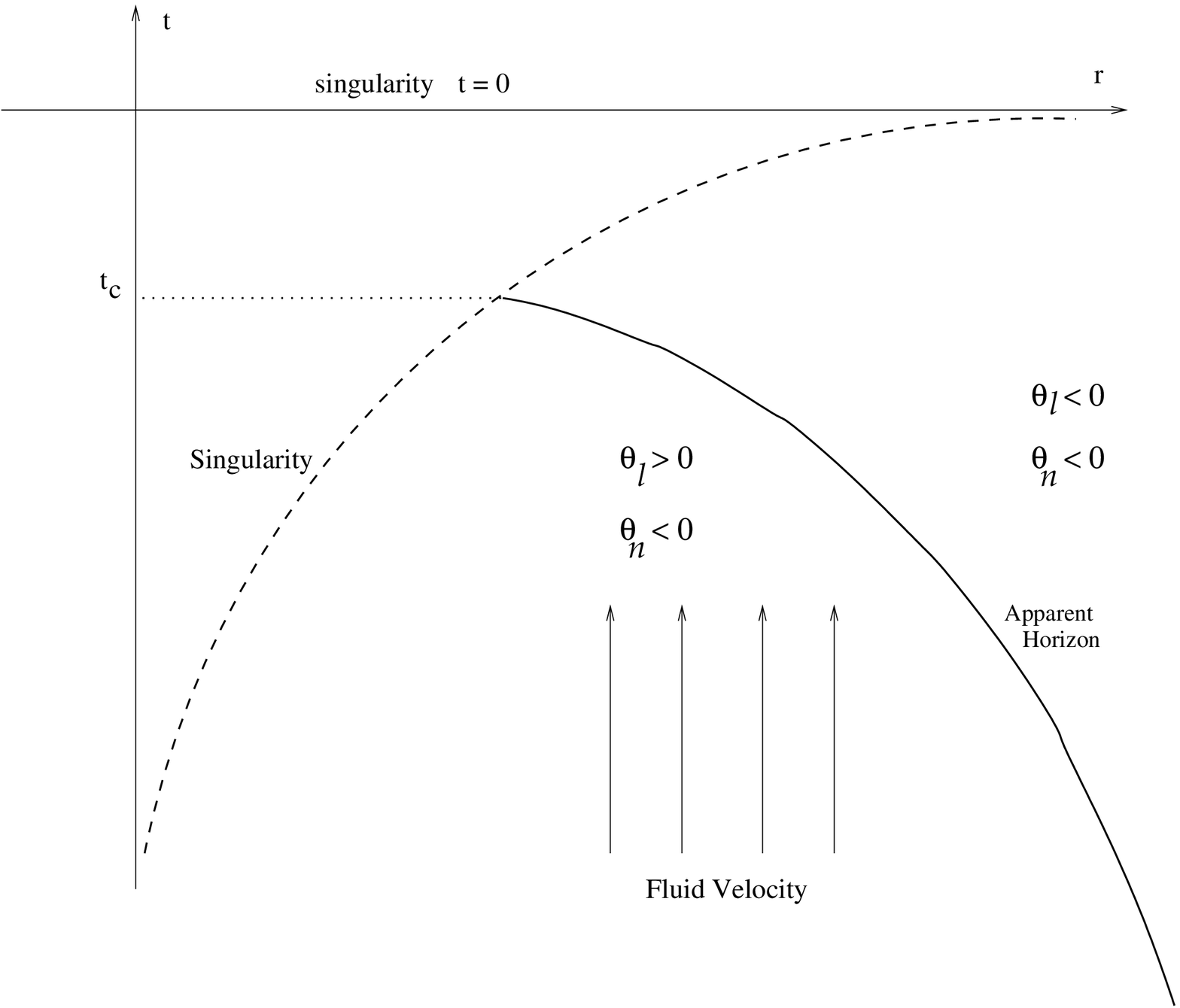}{This plot shows the main structures of the
 interior
spacetime,
before the matching with the exterior Schwarzschild spacetime. The
 dashed 
curve $(-t)=1/r^3$ represents the outest
singularity considered.}{singularity}

         \bq
         \lb{sing}
         (-t_{sing}) = \frac{1}{r^3 }  .
         \eq

From the Kretschmann scalar we can see that this singularity, $X=0$
(corresponding to $R=0$), 
represents a physical one, although it does not imply in the
divergence of the pressure. 

     For a perfect fluid, the energy conditions are given by the
 following
     expressions \cite{HE73}:

    a) weak energy condition:
     \bqn
     \lb{weak}
       \rho \geq 0 , \;\;\; \rho + p \ge 0  ,
     \eqn

     b) dominant energy condition:
     \bqn
     \lb{dominant}
      \rho + p \geq 0  , \;\;\;  \rho - p \geq 0  ,
         \eqn

      c) strong energy  condition:
    \bqn
     \lb{strong}
      \rho + 3p \geq 0  .
         \eqn

      For the energy density given by the equation
 (\ref{endensityperfluid}) and the
pressure given by equation (\ref{pressureperfluid}), these 
expressions take the forms:
      \bqn
      \lb{energyconditions}
      \rho + p = \frac{1}{3t^2} \left[ \left( \frac{2+3X}{1+3X} \right) \left( \frac{1+X}{X} \right) +1 \right]  ,  \nb\\
      \rho - p = \frac{1}{3t^2} \left[ \frac{2+4X}{(1+3X)X} \right]  , \nb\\
       \rho + 3p = \frac{1}{3t^2} \left[ \frac{2+8X+12X^2}{X(1+3X)} \right] ,
       \eqn
       which are all satisfied when the expression in brackets of
       the last formula is positive, i.e., for
        $(-t) \geq \frac{1}{r^3}$.

Note that by (\ref{endensityperfluid}), (\ref{pressureperfluid}) and 
(\ref{K}) we have that the pressure does not
diverge on the physical
singularity at $X=-1/3$, but the energy conditions diverge.

    The congruence of the outgoing and ingoing null geodesic expansion
    are given by \cite{Yasuda,Yasuda1,Yasuda2,Yasuda3}
    \bqn
     \lb{expansions2}
     \theta_l = \frac{f}{r S} \left[- \frac{S_0
 r^2}{3(-t)^{2/3}X^{1/3}}
     \left(\frac{2}{3} + X \right) + 1 \right]  \nb\\
     \theta_n = \frac{g}{r S} \left[-\frac{S_0 r^2}{3 (-t)^{2/3}
     X^{1/3}} \left(\frac{2}{3} + X \right) - 1 \right] ,
     \eqn
respectively.
Although we can not solve analytically the equation $\theta_l=0$ (which
defines the apparent horizon), we
 can find
graphically the curve $r_{AH}(t)$ shown in figure 1, which intercepts
 the
singularity at $t=t_c$.
The outgoing expansion is positive
outside of the apparent horizon and negative inside of it, while 
$\theta_n$ is always negative, characterizing the
structure of trapped surface. 
Note that for $t>t_c$ we have black hole, while for $t<t_c$ we get a
marginally naked singularity (see figure 1).
These results suggest that this solution can represent a scenario
where a naked singularity can be dressed by the collapse of a 
perfect fluid.

    \section{Matching with asymptotically flat spacetime}

     Since self-similar spacetimes are not asymptotically flat,
     then it is necessary to make a junction with a
     static and asymptotically flat spacetime, i.e., the
     Schwarzschild metric. 

      We use the coordinates $\tau, \theta, \phi$
     for the hypersurface junction in a comoving
     framework.

     Thus, we can write now equation (\ref{metricX}) as the interior
 metric in the
     following form
     \bqn
     \lb{interiormetric}
     ds^2_- = dt^2 - 4 S_0^2 X^{4/3} e^{2X} \left(1 + \frac{1}{3X}
     \right)^2 dr^2 - r^2 S_0^2 X^{4/3} e^{2X} d\Omega^2
     \eqn
         and the Schwarzschild metric is
        \bqn
        \lb{exteriormetric}
               ds^2_+ = \left(1- \frac{2m}{\it{R}} \right) dT^2 -
 \left(1- \frac{2m}{\it{R}}
                \right) ^{-1} d\it{R}^2 - \it{R}^2 d \Omega^2   .
        \eqn

       We can describe the junction hypersurface in the interior and
       exterior coordinates by the following expressions
       \bqn
        \lb{hypersurface}
         H_- = r -r_{\Sigma} = 0 \;\;     \mbox{in} \;\;  \nu^-   \nb\\
         H_+ = \it{R} - \it{R}_{\Sigma}(t) = 0  \;\;  \mbox{in} \;\;  
 \nu^+ ,
        \eqn
        then the on the hypersurface, making $r = constant$, the
        interior and exterior metrics becomes
        \bqn
        \lb{metricsurface}
         ds_-^2 = dt^2 - r_{\Sigma}^2 S_0^2 X^{4/3} e^{2X} d\Omega^2 ,
  \\
         ds_+^2 = dT^2\left[ \left(1 - \frac{2m}{\it{R_{\Sigma}}}
 \right) - \left(1 - \frac{2m}{\it{R_{\Sigma}}}
         \right)^{-1} \left(\frac{d\it{R_{\Sigma}}}{dT} \right)^2
  \right] -
         \it{R_{\Sigma}}^2 d\Omega^2 .
         \eqn

Comparing the first fundamental form furnished by the above metrics, we
 have
         \bqn
         \lb{firstfundamental}
          R_{\Sigma} = r_{\Sigma}^2 S_0 (-t)^{1/3} [ln(r_{\Sigma}
 (-t)^{1/3})^{2/3}  ,     \nb\\
           dt^2 = d\tau^2 = dT^2 \left[ \left(1 -
 \frac{2m}{\it{R_{\Sigma}}}
           \right) - \left(1 - \frac{2m}{\it{R_{\Sigma}}}
           \right)^{-1} \left(\frac{d\it{R_{\Sigma}}}{dT} \right)^2
 \right] ,
          \eqn
            from which we obtain an expression for $\frac{d
            \it{R}}{dT}$, that is, considering the hypersurface
 (dropping the
subscript $\Sigma$)
            \bqn
            \lb{derivativeR}
               \frac{d \it{R}}{dT} = \frac{-1 \pm \sqrt{1 + 4 r^2
 S^2_{,t}}}{2r S_{,t}}
               \left( 1 - \frac{2m}{\it{R}}  \right)  .
             \eqn

The interior e exterior extrinsic curvature are given by
\bq
K^{-}_{\tau \tau}=0,
\eq
\bq
K^{-}_{\theta \theta}= \frac{9}{2} S_0 X^{2/3} e^X \left( 1 +
 \frac{1}{3X}
\right) \frac{r}{(X+3)^2},
\eq
\bq
K^{+}_{\tau \tau}= \frac{d^2 R}{d\tau^2} - \frac{m}{R^3}(2m- R)
\left( \frac{dT}{d\tau} \right)^2 + \frac{m}{R(2m-R)}
\left( \frac{dR}{d\tau} \right)^2,
\eq
\bq
K^{+}_{\theta \theta}= - \frac{(2m - R)} { \left(1- \frac{2m}{ R}
 \right)
- \left(\frac{dR}{dT} \right)^2  \left(1 - \frac{2m}{R} \right)^{-1}}.
\eq

From the continuity of $K_{\tau \tau}$ we obtain
\bq
\frac{d^2 R}{dT^2} = \frac{m}{R}\left[ \frac{2m-R}{R} - 
\frac{1}{2m-R}\left(\frac{d R}{dT}\right)^2 \right].
\label{derivativeR2}
\eq
Since the radius of hypersurface junction is diminishing with the time and
radius of Schwarzschild event horizon is constant, it is reasonable to
expect that the junction radius will cross the horizon some time, forming
a black hole.  We can note from equations (\ref{derivativeR}) and (\ref{derivativeR2})
that, when $R \rightarrow 2m$, we have $\frac{d R}{dT}=\frac{d^2 R}{dT^2}=0$.

In the following we show that the junction is also possible if we consider
a thin shell separating the two spacetimes.

In this case, the expressions for the energy
     density and the pressure on the shell \cite{Jaimepaper2} are,
 respectively
      \bqn
      \lb{energyshell}
      \sigma &=& - \left\{ [K_{\tau \tau}] - [K] \right\} ,  \nb\\
      \eta &=& - \frac{1}{R^2} \left\{ [K_{\theta \theta} ] + R^2 [K]
  \right\}
           \eqn
       where $[K]$ and  $[K_{ab}]$ are

       \bqn
       \lb{escalarextrinsic1}
              [K] &=& g^{ab} [K_{ab}] , \nb\\
       \left[K_{ab}\right] &=& {K^+}_{ab} - {K^-}_{ab}
         \eqn
         and $K_{ab}$ is the extrinsic curvature, given by

         \bqn
         \lb{extrinsiccurvature}
         K_{ab} = - n^i \left[ \frac{\partial^2 x^i}{\partial \xi^a
 \partial \xi^b}
             + \Gamma^i_{jk}  \frac{\partial x^j}{\partial \xi^a}
             \frac{\partial x^k}{\partial \xi^b} \right] ,
        \eqn
      with the energy momentum tensor of the shell given by the
 following expression
        \bqn
    \lb{tensorshell}
    S_{ab} = -  \left\{ K_{ab} - g_{ab} [K] \right\} .
    \eqn

     Using the solution (\ref{solperfectfluid}) and the expressions for
 the energy density and the
      pressure for the shell
       (\ref{energyshell}), we obtain
         \bqn
     \lb{densityshell}
     \sigma &=&  \left\{ \frac{2}{r^2 S_0^2 X^{4/3} e^{2X}}
   \frac{2m - \it{R}}{ \left[ \left(1- \frac{2m}{\it{R}} \right) -
 \left(\frac{d\it{R}}{dT}\right)^2
    \left(1- \frac{2m}{\it{R}} \right)^{-1}  \right] } \right.    \nb\\
       &+& \left. \frac{9}{r S_0  X^{2/3} e^X (3+X)^2}
    \left(1 + \frac{1}{3X} \right)  \right\}   ,
         \eqn
           and
     \bqn
       \lb{pressureshell}
     \eta &=&  \left\{ \frac{1}{\left(1 - \frac{2m}{\it{R}}\right) -
 \left(\frac{d\it{R}}{dT}\right)^2
       \left(1 - \frac{2m}{\it{R}}\right)^{-1}} \left[
 \frac{d\it{R}}{dT}
       \left[- \frac{\partial^2 T}{\partial \tau^2}
             + \frac{2m}{\it{R}(\it{R} +
       2m)} \frac{\partial T}{\partial \tau} \frac{\partial
 \it{R}}{\partial \tau}
       \right]                  \right. \right.   \nb\\
          &+& \left. \left[ \frac{\partial^2 \it{R}}{\partial \tau^2}
  - \frac{m(2m -
          \it{R})}{r^3} \left(\frac{\partial T}{\partial \tau}
          \right)^2
           + \frac{m}{\it{R}(2m - \it{R})} \left( \frac{\partial
          \it{R}}{\partial \tau}\right)^2 \right] \right]            
 \nb\\
          &-&  \frac{1}{r^2 S_0^2 X^{4/3} e^{2X} } \left[ \frac{(2m -
 \it{R})}{\left(1- \frac{2m}{\it{R} } \right)
           -\left(\frac{d \it{R}}{dT} \right)^2 \left(1-
           \frac{2m}{\it{R}}
           \right)^{-1} }          \right.         \nb\\
           &+&\left. \left.  \frac{9}{2} S_0 X^{2/3} e^X \left(1 +
           \frac{1}{3X} \right) \frac{r}{(3+ X)^2} \right] \right\}
           .
           \eqn

            We require for the energy tensor of the shell that the
            energy density must be positive, then
            we should have
            \bqn
             \lb{conditionshell}
            -  \left[ 1 - \left( \frac{-1 \pm \sqrt{1 + 4 r^2
 S^2_{,t}}}{2r S_{,t}} \right)^2\right]^{-1}
              + \frac{9}{2} \left( 1 + \frac{1}{3X} \right)
              \frac{1}{(3+ X)^2}  > 0 ,
             \eqn
              which means, if we choose the minus sign in the
              first term of the above expression, it is always
              positive. Thus, it was possible to do the matching with
               a shell and the exterior spacetime is Schwarzschild
               one, characterizing a black hole as the final structure 
of the collapse process.

\section{Conclusion}

In this work we have studied the evolution of a perfect fluid which
collapses into an initial naked singularity.

We have used the self-similar general solution of the
Einstein field equations, for $\alpha=-3$, by assuming that the fluid moves along
time-like geodesics. The energy conditions, geometrical and physical
properties of the solution was studied.

The analysis of the congruence of the outgoing and ingoing null geodesic expansion
have shown that the initial naked singularity evolve to a black hole after a 
particular time $t=t_c$.  This can be clearly seen in figure 1, where the spacetime
around the singularity is initially untrapped ($\theta_l > 0$, $\theta_n < 0$) and
after $t=t_c$ it has become trapped ($\theta_l < 0$, $\theta_n < 0$).
 
In order to be sure that the final structure is really a black hole,
we have shown that the matching with the asymptotically and static spacetime,
the Schwarzschild one, is possible, either considering a thin shell
or not.

We have presented an example where an initial naked singularity
can be dressed and it becomes a black hole by collapsing of standard matter.
A similar result was shown by Brandt et al. \cite{rev7}.

Finally, this result and the fact that the
pressure does not diverge on the naked singularity suggests that there can
exist a connection between naked singularities and some kind of
weakness of the gravitational field, compared to that associated
to black holes.

\section*{Acknowledgments}

The financial assistance from 
FAPERJ/UERJ (MFAdaS, CFCB) is gratefully acknowledged.
The author (R.C.) acknowledges the financial  support from FAPERJ.
The authors (RC and MFAS) also acknowledge the financial support from
Conselho Nacional de Desenvolvimento Cient\'{\i}fico e Tecnol\'ogico -
Brazil.

\label{lastpage}


\begin{thebibliography}{100}


\bibitem{rev1} R. Penrose, in {\it Black holes and relativistic stars}, ed. R. M. Wald, University of Chicago Press (1998). 

\bibitem{rev2} A. Krolak, Prog. Theor. Phys. Suppl. {\bf 136}, 45 (1999).

\bibitem{rev3} P. S. Joshi, Pramana {\bf 55}, 529 (2000). 

\bibitem{rev4} M. Celerier and P. Szekeres, Phys.Rev. D {\bf 65}, 123516 (2002). 

\bibitem{rev5} R. Giambo, F. Giannoni, G. Magli, P.  Piccione, Commun. Math. Phys. {\bf 235}, 545 (2003). 

\bibitem{rev6} T. Harada, H. Iguchi, and K. Nakao, Prog.Theor.Phys. {\bf 107}, 449 (2002).

\bibitem{rev7} Brandt, C.F.C., Lin, L.-M., Villas da Rocha, J.F., Wang, A.Z., Int. J. Mod. Phys. D, {\bf 11}, 155 (2002).

\bibitem{rev8} Chan, R., da Silva, M.F.A., Villas da Rocha, J.F., Int. J. Mod. Phys. D, {\bf 12}, 347 (2003).

\bibitem{dust1} D. M. Eardley and L. Smarr, Phys. Rev. D {\bf 19}, 2239 (1979). 

\bibitem{dust2} D. Christodoulou, Commun. Math. Phys., {\bf 93}, 171 (1984). 

\bibitem{dust3} R. P. A. C. Newman, Class. Quantum Grav., {\bf 3}, 527 (1986).

\bibitem{jd} P. S. Joshi and I. H. Dwivedi, Phys. Rev. D {\bf 47}, 5357 (1993).

\bibitem{HE73} S.W. Hawking and G.F.R. Ellis, {\em The Large Scale
Structure of Spacetime}, Cambridge University Press, Cambridge
(1973).

\bibitem{Yasuda} A.Y. Miguelote, N.A. Tomimura, and A. Wang, Gen.
 Relativ.
Grav. {\bf 36}, 1883 (2004) [arXiv:gr-qc/0304035].

\bibitem{Yasuda1} A. Wang, Phys. Rev. D {\bf 72}, 108501 (2005)
[arXiv:gr-qc/0309003].

\bibitem{Yasuda2} A. Wang, Phys. Rev. {\bf D68} 064006 (2003)
[arXiv:gr-qc/0307071].

\bibitem{Yasuda3} A. Wang, Gen. Relativ. Grav. {\bf 37}, 1919 (2005)
[arXiv:gr-qc/0309005].

\bibitem{Brandt2003}
C.F.C. Brandt, M.F.A. da Silva and J.F. Villas da Rocha, R. Chan,
Int. J. Mod. Phys. D, {\bf 12}, 1315 (2003).

\bibitem{Jaimepaper2} J.F. Villas da Rocha,  A.Z. Wang, N.O. Santos,
Phys. Lett. {\bf A 255}, 213 (1999).

\end{thebibliography}
\end{document}